\documentclass[eqsecnum,showpacs,aps,epsf,elsart,timesmm]{revtex4}
\usepackage{amsfonts,latexsym,eucal,amsmath,amssymb,times,mathrsfs
}
\usepackage[usenames]{color}
\newcommand{\bk}{{\mbox{\boldmath{\scriptsize $k$}}}}
\newcommand{\bK}{{\mbox{\boldmath{\scriptsize $K$}}}}
\newcommand{\bp}{{\mbox{\boldmath{\scriptsize $p$}}}}
\newcommand{\bm}[1]{\hbox{\boldmath{$#1$}}}
\newcommand{\sbm}[1]{\hbox{\boldmath{\scriptsize$#1$}}}
\newcommand{\Bk}{{\mbox{\boldmath{$k$}}}}
\newcommand{\BK}{{\mbox{\boldmath{$K$}}}}
\newcommand{\Bp}{{\mbox{\boldmath{$p$}}}}

\newcommand{\Bx}{{\mbox{\boldmath{$x$}}}}
\newcommand{\bX}{{\mbox{\boldmath{\scriptsize $X$}}}}
\newcommand{\BX}{{\mbox{\boldmath{$X$}}}}
\newcommand{\Bpartial}{{\mbox{\boldmath{$\partial$}}}}

\newcommand{\dd}{{\rm d}}
\newcommand{\Approx}{\stackrel{\rm IR}{\approx}}
\newcommand{\cd}{\cdot}
\newcommand{\Mp}{M_{\rm pl}}
\newcommand{\sR}{{^s\!R}}
\newcommand{\gz}{{^g\!\zeta}}

\begin{document}
\title{Dominance of gauge artifact in the consistency relation for the
primordial bispectrum}
\author{Takahiro Tanaka$^{1}$}
\email{tanaka_at_yukawa.kyoto-u.ac.jp}
\author{Yuko Urakawa$^{2,3}$}
\email{yurakawa_at_ffn.ub.es}
\address{
$^{1}$ Yukawa Institute for Theoretical Physics, Kyoto university, Kyoto, 606-8502, Japan\\
$^{2}$ Departament de F{\'\i}sica Fonamental i Institut de Ci{\`e}ncies del Cosmos, 
Universitat de Barcelona,
Mart{\'\i}\ i Franqu{\`e}s 1, 08028 Barcelona, Spain\\
$^{3}$ Department of Physics, Ochanomizu University 2-1-1 Otsuka, Bunkyo, Tokyo, 112-8610 Japan  }
\begin{abstract}
The conventional cosmological perturbation theory has been performed under the assumption that 
we know the whole spatial region of the universe with infinite
 volume. This is, however, not the case in the actual observations
 because observable portion of the universe is limited. 
To give a theoretical prediction to the observable
fluctuations, gauge-invariant observables should be composed of 
the information in our local observable universe with finite volume. 
From this point of view, we reexamine the primordial non-Gaussianity in single
field models, focusing on the bispectrum in the squeezed limit. 
A conventional prediction states that the bispectrum in this limit 
is related to the power spectrum through the so-called consistency
relation. However, it turns out that, if we adopt a genuine gauge invariant 
variable which is naturally composed purely of the information in 
our local universe, the leading term for the
 bispectrum in the squeezed limit predicted by the consistency relation
 vanishes. 
\end{abstract}
\pacs{98.80.Cq, 98.80.Es, 98.80.Jk}
\maketitle

\section{Introduction}

The measurements of the cosmological fluctuations have now become a 
crucial tool to probe the history of the early universe. To yield the
theoretical prediction of the fluctuations, which are to 
be compared with the observations, the gauge-invariant
perturbation theory is used. 
It is not, however, widely recognized that the gauge invariance in the whole
universe with infinite volume should be distinguished from that 
in the local universe
with finite volume. 
It is widely known that the gauge-invariant cosmological 
perturbation can be realized by completely fixing the coordinates.
In the case of the gauge invariance in the whole universe, where we
request the invariance under normalizable gauge transformations,
the conventional gauge fixing conditions are sufficient to uniquely 
determine the coordinates. 
By contrast, if we require that the coordinates within our local observable universe should be 
completely fixed only from the local information contained therein, the conventional gauge conditions 
are not sufficient. To fix the gauge locally, these conditions
must be supplemented with appropriate boundary conditions~\cite{IRsingle}.  
These boundary conditions are fixed by requiring the
regularity at the spatial infinity in the conventional approach.
We should remind that, in the actual observations such as measurements of the
Cosmic Microwave Background (CMB), we can observe only our local
universe and we cannot consult the regularity at infinity. 
Therefore, the observable fluctuations should be calculated in the
request of the invariance under both the normalizable and non-normalizable
gauge transformations.
We distinguish gauge-invariant quantities that can be constructed from our local 
observable universe, referring them as {\it genuine gauge-invariant variables}.
The observable fluctuations should be such a genuinely gauge-invariant variable.  
 
The importance of the genuine gauge invariance is
also highlighted 
in the context of the infrared~(IR) divergence
problem. The adiabatic vacuum, which yields the scale-invariant spectrum,
is supposed to be a natural vacuum in the inflationary
universe. However, once the interaction turns on, it is not manifest whether
the adiabatic vacuum is stable or unstable against the IR contributions in loop
corrections. It is indeed known that the loop corrections calculated in
the conventional perturbation theory yield
the logarithmic divergences~\cite{Boyanovsky:2004gq, Boyanovsky:2004ph, Boyanovsky:2005sh,
Boyanovsky:2005px, Onemli:2002hr, Brunier:2004sb, Prokopec:2007ak,
Sloth:2006az, Sloth:2006nu, Seery:2007we, Seery:2007wf, Urakawa:2008rb,
Adshead:2008gk, Cogollo:2008bi, Rodriguez:2008hy, Seery:2009hs,
Gao:2009fx, Bartolo:2010bu, Seery:2010kh,
Kahya:2010xh}. Our previous works~\cite{IRgauge_L, IRgauge}
indicate that this divergence is an unphysical artifact 
in the case of single field models of inflation. 
We found that the initial quantum states should satisfy certain 
conditions if we require that the loop correction to the two-point 
function of a genuine gauge invariant variable should be free from the IR divergence. 
Our interpretation is that these conditions are the requirements 
that the initial states are invariant under the residual gauge
transformation. 
Namely, as long as the 
initial quantum states respect the gauge invariance in our 
local universe, the IR contribution to the loop correction 
should be suppressed. Hence, we referred to these conditions on the 
initial quantum states as the {\em gauge invariance conditions}. 

In this paper we reexamine the primordial non-Gaussianity in single
field models, taking into account the importance of genuine gauge
invariance of the selected variables and the gauge invariance 
conditions on the initial quantum state. 
A conventional prediction states that the bispectrum 
in the squeezed limit is related to the amplitude and the spectral 
index of the power spectrum, which is often called the consistency
relation. However, it will turn out that the bispectrum of a genuine gauge 
invariant variable, which is the counterpart of the conventional 
gauge invariant curvature perturbation, is more suppressed 
in the squeezed limit compared with the prediction widely accepted 
as the consistency relation. 

This paper is organized as follows. In Sec.~\ref{Sec:GILocal}, we 
review a method to introduce 
genuinely gauge-invariant variables and 
the gauge invariance conditions. 
The basic perturbation equations will be also provided there. 
In Sec.~\ref{Sec:NG}, for illustrative
purpose, we first briefly summarize the usual argument for 
the consistency relation, which uses the information 
about the region outside our observable universe. 
After that, we re-investigate the bispectrum in the squeezed limit, 
using a genuinely gauge-invariant variable and 
imposing the gauge invariance conditions on the initial quantum state. 
Section~\ref{Sec:Conclusion} is devoted to the conclusion.

\section{Gauge-invariant perturbations in local universe} \label{Sec:GILocal}

Gauge invariance is guaranteed if we compute after 
complete gauge fixing. However, complete gauge fixing is 
not so straightforward as mentioned in the introduction. 
Let's first impose conventional comoving gauge conditions 
\begin{eqnarray}
 \delta \phi=0\,,
\label{eq1} 
\end{eqnarray}
and
\begin{eqnarray}
 \delta {\gamma^i}_{i}=0=\partial_i \delta {\gamma^i}_j\,,  
\label{eq2} 
\end{eqnarray}
with 
\begin{equation}
h_{ij}= e^{2(\rho+\zeta)} \left[ e^{\delta\gamma}
 \right]_{ij}\,, 
\end{equation}
where we denote the scale factor by $e^\rho$. The time coordinate is then fixed completely 
by the first condition~(\ref{eq1}). The second conditions~(\ref{eq2}), however, do not fix the 
spatial coordinates completely if we just consider 
the local universe and adopt gauge conditions only in this region.
Namely, there are still residual gauge 
degrees of freedom as long as only the local metric is concerned.
If we could use the information stored in the region outside our
observable universe to fix the coordinates, these residual gauge degrees
of freedom would disappear.
In spite of this, once we accept the contamination from the 
outside of our observable universe to 
define gauge invariant variables, 
such quantities cannot correspond to actual observables in a 
strict sense. 
Furthermore, such contamination from outside of our observable universe 
may cause artificial infrared divergences. 
We therefore need to fix the gauge completely using only 
the quantities residing in our observable region, but this is technically complicated. 

In our previous paper we pursued an alternative 
way to introduce genuine gauge-invariant variables, 
focusing on the scalar 
under three-dimensional diffeomorphism, e.g. three-curvature on 
the time-constant hypersurface. Such a quantity should be 
a genuine gauge invariant if the spatial position is specified 
in a coordinate independent manner. In order to specify 
a point in a coordinate independent manner, we used the geodesic 
normal coordinates spun from a representative observer. 
There remains the ambiguity in choosing the position of 
this representative observer and his/her frame. 
This dependence, however, should disappear
in the evaluation of $n$-point correlation functions as long as 
the initial quantum state respects the translational 
and rotational symmetries. 

We computed the two-point correlation function of thus
defined genuine gauge-invariant operator at one loop level, focusing on the effect of IR modes. 
Although the IR contribution to the two-point function 
is divergent in general if the spectrum of the linear 
curvature perturbation is scale invariant or red, we found that, 
if we choose the initial quantum states appropriately, 
this divergences disappear. We derived the conditions for the 
absence of IR divergences and confirmed that such quantum states 
exist at least up to the second order in the slow roll expansion. 

Now it is profitable to ask the question how to interpret the divergences
for generic quantum states. 
Since Fourier modes of $\zeta$ basically become constant in the IR limit,
they can be almost absorbed by the overall 
scale transformation of the spatial coordinates, 
which is a part of the residual gauge degrees of freedom mentioned above. 
Although it is just a gauge, the fluctuation 
of this mode causes infinitely large variance of 
the two-point function in conventional computations. Hence, it was speculated that this divergence 
occurs because generic initial quantum states do not respect the 
expected invariance under these non-normalizable gauge transformations. 
Taking these in mind, we refer to the conditions necessary to avoid IR
divergences in $n$-point functions as the gauge invariance conditions. 
In this section, we briefly 
review these gauge invariance conditions
derived in our previous works~\cite{IRgauge_L, IRgauge}.

\subsection{Gauge-invariant operator} \label{SSec:GIQ}

The construction of genuine gauge-invariant quantities can be
achieved simply by making use of scalar quantities under 
three-dimensional spatial diffeomorphism. 
In our previous works~\cite{IRgauge_L,
IRgauge}, we picked up the scalar 
curvature $\sR$ as such a scalar quantity. 
Although scalar quantities also vary under
the gauge transformation because of the change of their arguments
$x^i$, this gauge ambiguity does not appear in the $n$-point functions
of these quantities if the arguments are specified in a gauge-invariant
manner. In order to specify the arguments of the $n$-point functions
in a gauge invariant manner, we measure distances from an
arbitrary reference point to the $n$ vertices by means of 
the geodesic distance obtained by solving the spatial three-dimensional 
geodesic equation: 
\begin{eqnarray}
 \frac{\dd^2 x^i}{\dd \lambda^2} +  {^s \Gamma^i}_{jk} \frac{\dd
  x^j}{\dd \lambda} \frac{\dd x^k}{\dd \lambda} =0~,
\label{GE}
\end{eqnarray}
where ${^s \Gamma^i}_{jk}$ is the Christoffel symbol with respect to 
the three dimensional spatial metric on a constant time hypersurface and
$\lambda$ is the affine parameter. 
We consider the three-dimensional geodesics whose affine parameter ranges
from $\lambda=0$ to $1$ with the initial ``velocity'' given by
\begin{eqnarray}
 \left.{\dd x^i(\bm{X},\lambda)
\over \dd \lambda}\right\vert_{\lambda=0}= e^{-\zeta(\lambda=0)}\! \left[
e^{-\delta \gamma(\lambda=0)/2}\right]\!^i\!_j\, 
X^{j}\,. \label{IC}
\end{eqnarray}
We identify a point in the geodesic normal coordinates $X^i$ 
with the end point of the geodesic, $x^i(\bm{X},\lambda=1)$.

During inflation, the scales relevant to the current observations go
far outside of the inflationary horizon, which leads to the natural assumption that
the observable scale is much larger than the horizon
scale $1/H$. In Sec.~\ref{Sec:NG} we will evaluate the fluctuation
with the modes $1/L_{\rm obs} \ll k \ll e^\rho H$, where
$L_{\rm obs}$ denotes the observable scale in the comoving
coordinates. In this section, however,
it is sufficient to restrict the expression for $x^i$ to the case
with $1/L_{\rm obs} \gg k$, as we are interested in the IR divergence.
In this case ${^s\Gamma^i}_{jk}$ is suppressed since it contains a spatial
differentiation. Then, the geodesic normal coordinates $X^i$ are
approximately related to the global coordinates $x^i$ as
\begin{align}
 & x^i(\bm{X})=:X^i+\delta x^i(\bm{X}) \simeq e^{-\zeta}\!
\left[ e^{-\delta \gamma/2} \right]^i\!_j\, X^j\,.
\label{Exp:Xi}
\end{align}
We use this simple expression in this section, and we defer its
generalization to the case with $1/L_{\rm obs} \ll k$ to
Sec.~\ref{SSec:GI}.

Using the geodesic normal coordinates $X^i$, a genuine gauge invariant
variable can be constructed as
\begin{align}
 & {^g\!R}(X) := \sR (t,\, x^i(\bm{X})) = \sum_{n=0}^\infty \frac{\delta x^{i_1} \cdots \delta x^{i_n}}{n!}
 \partial_{i_1} \cdots \partial_{i_n}\!{^s\!R}(t,\,
 x^i)\vert_{x^i=X^i} \,, \label{Def:gR}
\end{align}
where $t$ denotes the cosmological time and we introduced the
abbreviated notation $X:=\{t,\,\bm{X}\}$.

\subsection{Solving non-linear perturbation} 
In this paper we consider a single scalar field with the canonical
kinetic term, whose action takes the form 
\begin{eqnarray}
 S = \frac{\Mp^2}{2} \int \sqrt{-g}~ [R - g^{\mu\nu}\phi_{,\mu} \phi_{,\nu} 
   - 2 V(\phi) ] \dd^4x~,
\end{eqnarray}
where $\Mp$ is the Planck mass and the scalar field $\phi$ was 
rescaled as $\phi \to \phi/\Mp$ to be dimensionless. 
We use the ADM formalism,  
following Ref.~\cite{Maldacena}. 
The metric in the ADM form is given by 
\begin{eqnarray}
 \dd s^2 = - N^2 \dd t^2  + h_{ij} (\dd x^i + N^i \dd t) (\dd x^j + N^j
  \dd t)~. 
\end{eqnarray}
The above action is then rewritten as
\begin{eqnarray}
 S&\!=&\!\frac{\Mp^2}{2} \int\! \sqrt{h} \Bigl[ N \,\sR - 2 N
  V(\phi) + \frac{1}{N} (E_{ij} E^{ij} - E^2) 
 + \frac{1}{N} ( \partial_t \phi
  - N^i \partial_i \phi )^2 - N h^{ij} \partial_i \phi \partial_j \phi
  \Bigr] \dd^4x~, 
\end{eqnarray}
where $\sR$ is the three-dimensional scalar curvature and $E_{ij}$ and $E$ are defined by 
\begin{eqnarray}
 E_{ij} = \frac{1}{2} \left( \partial_t h_{ij} - D_i N_j
 - D_j N_i \right), \qquad \quad E = h^{ij} E_{ij} ~.
\end{eqnarray}
Here $D_i$ denotes the three-dimensional covariant
derivative and the spatial indices are raised and lowered by $h_{ij}$.
To calculate the non-linear corrections under the slow-roll
approximation, it is convenient to temporally 
work in the flat gauge: 
\begin{eqnarray}
 \tilde{h}_{ij}= e^{2\rho} \left[ e^{\delta \tilde\gamma} \right]_{ij}\,,
\label{Exp:metricflat}
\end{eqnarray}
with the transverse-traceless conditions 
\begin{eqnarray}
   {\delta \tilde{\gamma}^i}_{i}=0=
  \partial_i {\delta \tilde{\gamma}^i}_{j}\,, 
\end{eqnarray}
because all the interaction vertices are explicitly suppressed by the
slow-roll parameters in this gauge~\cite{Maldacena, IRsingle}. 
Here in this section we associate a tilde with the
metric perturbations in the flat gauge to discriminate from 
those in the comoving gauge. The action in this gauge is given by
\begin{align}
 S \Approx 
 \frac{\Mp^2}{2} \!\int\! e^{3\rho} \Bigl[\tilde{N}^{-1}\!
 \left(-6\dot{\rho}^2+4\dot{\rho} \partial_i \tilde{N}^i \right)+  \tilde{N}^{-1} 
  \left( \dot\phi + \dot\varphi - \tilde{N}^i \partial_i \varphi
 \right)^2 
 - 2 \tilde{N} \!\sum_{m=0} \frac{V^{(m)}}{m!}\varphi^m
 - \tilde{N} \tilde{h}^{ij} \partial_i \varphi
 \partial_j \varphi  \Bigr]\! \dd t\, \dd^3 \bm{x}\,,
 \label{Exp:S/flat}
\end{align}
where $V^{(m)}:= \dd^m V/ \dd \phi^m$. 
We use the symbol ``$\Approx$'' as in \cite{IRgauge} to denote an equality
which is valid when we neglect the terms that do not participate in the IR
divergences. When we write down the 
Heisenberg operator in terms of the interaction picture field, 
this is equivalent to keep only terms without spatial and/or time
derivatives and the terms that include only one interaction picture 
field with differentiation.

We also adopt the 
slow-roll approximation for the background evolution. 
To characterize the deviation from the exact de
Sitter evolution, we use the horizon flow functions: 
\begin{eqnarray}
 \varepsilon_0 := \frac{H_i}{H}~, \quad
 \varepsilon_{m+1} := \frac{1}{\varepsilon_m} \frac{\dd
 \varepsilon_m}{\dd \rho}~\qquad \quad {\rm for}~m \geq 0,  \label{Def:HFF}
\end{eqnarray}
where $H$ is the Hubble parameter and $H_i$ is its value 
at the initial time.
The horizon flow functions are related to the conventional slow-roll
parameters as shown in Ref.~\cite{Schwarz:2001vv}. Hereafter, assuming that the
horizon flow functions $\varepsilon_m$ with $m \geq 1$ are
all small of ${\cal O}(\varepsilon)$, we neglect 
the terms of ${\cal O}(\varepsilon^3)$.

As studied in Appendix of Ref.~\cite{Maldacena}, the curvature perturbation in 
the comoving gauge $\zeta$
is related to the fluctuation of the dimensionless scalar field (divided
by $\Mp$) in the flat gauge $\varphi$ as
\begin{eqnarray}
 \zeta \Approx \zeta_n + \frac{1}{4} \varepsilon_2
  \zeta_n^2  + \zeta_n \partial_\rho \zeta_n \,, \label{Exp:zeta}
\end{eqnarray}
where we have introduced $\zeta_n:=- (\dot\rho/ \dot\phi) \varphi$ as in
Ref.~\cite{Maldacena}. 
In the calculation of the $n$-point functions, we solve the evolution
equation (Heisenberg equation) for the operator $\varphi$ 
iteratively to express $\varphi$ in terms of the interaction picture
field for $\varphi$. Variation of the total action with respect to $\varphi$
yields 
\begin{align}
 & e^{-3\rho}  \partial_t
 \left[ \frac{e^{3\rho}}{\tilde{N}} \left(\dot\phi + \dot\varphi \right)
 \right] + \tilde{N}  \sum_{m=0}^2 \frac{V^{(m+1)}}{m!}
 \varphi^{m} 
 - \left( \dot\phi + \dot\varphi \right) \frac{1}{\tilde{N}}
 \partial_i \tilde{N}^i - \tilde{N}e^{-2\rho}  \left[ e^{-\delta \tilde{\gamma}}
			       \right]^{ij} 
 \partial_i \partial_j \varphi  \Approx 0 ~.  \label{Eq:flat}
\end{align}
Variations with
respect to the lapse function and the shift vector, respectively, yield the
Hamiltonian constraint:
\begin{align}
 (\tilde{N}^2-1)  V  +  \tilde{N}^2  \sum_{m=1}^2
 \frac{V^{(m)}}{m!} \varphi^m 
  + 2 \dot\rho \partial_i \tilde{N}^i +  \dot\phi \dot\varphi+
 \frac{1}{2} {\dot\varphi}^2 \Approx 0\,,
\end{align}
and the momentum constraints:
\begin{eqnarray}
 2\dot\rho \partial_i \tilde{N} - \tilde{N}(\dot\phi \partial_i \varphi + \partial_i \varphi
  \dot\varphi) \Approx 0\,.
\end{eqnarray}
 These constraint equations are solved to give
\begin{eqnarray}
  \delta \tilde{N}& \Approx&  -{{\dot\phi}^2\over
   2{\dot\rho}^2}\zeta_n+{1\over 4 \dot\rho}\varphi \left(
   \dot\phi \delta \tilde{N}_1 + \dot\varphi \right)
   \Approx  -\varepsilon_1 \zeta_n+ \frac{\varepsilon_1}{2} \left(
							 \varepsilon_1 +
							\frac{\varepsilon_2}{2}\right)
  \zeta_{n}^2, \label{Exp:N/flat} \\ 
   \partial_i \tilde{N}^i& \Approx& \varepsilon_1 \dot\zeta_n
  -\frac{1}{2} \varepsilon_1 \varepsilon_2 \zeta_n \dot\zeta_n. \label{Exp:Ni/flat}
\end{eqnarray}
Substituting Eqs.~(\ref{Exp:N/flat}) and (\ref{Exp:Ni/flat}) into
Eq.~(\ref{Eq:flat}), the evolution equation of $\zeta_n$ is recast into 
a rather compact expression, 
\begin{eqnarray}
 {\cal L} \zeta_n&\Approx&  
 -2\varepsilon_1 \zeta_n  \frac{e^{-2\rho}}{{\dot\rho}^2}
   \partial^2 \zeta_n 
-  \varepsilon_1 \varepsilon_2  \zeta_n
  \partial_\rho \zeta_n - \frac{3}{4}  \varepsilon_2
  \varepsilon_3  \zeta_n^2 
  - \delta \tilde{\gamma}^{ij} \frac{e^{-2\rho}}{{\dot\rho}^2}  \partial_i \partial_j \zeta_n\,,
  \label{Eq:flat2}
\end{eqnarray} 
where, changing the time coordinate from $t$ to $\rho$, the differential operator ${\cal L}$ is given by 
\begin{equation}
{\cal L} :=  \partial_\rho^2 +  (3 - \varepsilon_1+ \varepsilon_2) \partial_\rho
 -  \frac{e^{-2\rho}}{{\dot\rho}^2} \partial^2\,.
\end{equation}
Expanding $\zeta_n$ and $\delta \tilde{\gamma}_{ij}$ as 
$$
\zeta_n= \psi + \zeta_{n,2} + \cdots, \qquad
\delta \tilde{\gamma}_{ij}= \delta \tilde{\gamma}_{ij,1} + \delta
\tilde{\gamma}_{ij,2} + \cdots,
$$ 
the equation of motion (\ref{Eq:flat2}) is reduced to 
\begin{align}
 & \quad {\cal L} \psi = 0~, \label{Eq:zetan1} \\
 & {\cal L} \zeta_{n,2} \Approx - \varepsilon_1 \varepsilon_2 \psi
 \partial_\rho \psi - \frac{3}{4} \varepsilon_2 \varepsilon_3 \psi^2 
 - 2 \varepsilon_1 \psi  \frac{e^{-2\rho}}{{\dot\rho}^2} \partial^2 \psi
 - \frac{e^{-2\rho}}{{\dot\rho}^2} \delta \tilde{\gamma}_1^{ij} \partial_i \partial_j \psi  ~. \label{Eq:zetan2} 
\end{align}
The first-order equation (\ref{Eq:zetan1}) gives the mode equation for
the interaction picture field and the second-order equation (\ref{Eq:zetan2}) is
integrated to give
\begin{align}
 \zeta_{n,2}(X)& \Approx \left( \frac{\varepsilon_1}{2} + \xi_2 \right) \psi^2
   + \varepsilon_1  \psi \partial_\rho\psi 
   + \varepsilon_1 (\varepsilon_1 + \varepsilon_2) \psi \partial_\rho
 \psi 
+\delta\zeta_{n,2} + \lambda_2
 \psi(\partial_\rho -X^i \partial_{X^i}) \psi + \frac{1}{2} \delta
 \tilde{\gamma}_1^{ij} X_i \partial_{X^j} \psi\,. \label{Sol:zetan2}
\end{align}
Here, $\zeta_{n,2}$ includes the non-local term:
\begin{equation}
  \delta\zeta_{n,2}:=-{\cal L}^{-1}
  \left[ \frac{3}{4}  \varepsilon_2 (2 \varepsilon_1 +
   \varepsilon_3) \psi^2 \right]~. 
\end{equation}
We kept the terms with second derivatives on the right hand side of
Eq.~(\ref{Eq:zetan2}) because they are also necessary to keep the terms in
$\zeta_{n,2}$ which have only one interaction picture field $\psi$ with differentiation.
It should be emphasized that the homogeneous solutions 
$\xi_2 \psi^2$ and $\lambda_2 \psi(\partial_\rho -X^i \partial_{X^i}) \psi$,  
in the sense that they satisfy ${\cal L}\cdots \Approx {\cal O}(\epsilon^3)$, 
can be added to
$\breve{\zeta}_{n,2}$, where the time dependent functions $\xi_2$ and $\lambda_2$
should be ${\cal O}(\varepsilon^2)$ and their derivatives should be 
${\cal O}(\varepsilon^3)$.

\subsection{Gauge invariance conditions of initial state} 

Now, we address the gauge-invariance conditions to be imposed on the
initial state, where initial conditions have two aspects. 
One is how we choose the positive frequency function for 
the interaction picture field $\psi$ in solving the mode
equation (\ref{Eq:zetan1}). The other is how we construct 
the non-linear Heisenberg field iteratively from the 
interaction picture field, i.e. how we add homogeneous solutions in
Eq.~(\ref{Sol:zetan2}). The interaction picture field $\psi$ is quantized as
\begin{eqnarray}
 \psi(X) = \int \frac{\dd^3 \bm{k}}{(2\pi)^{3/2}} e^{i
  \bk \cdot \sbm{X}} \psi_{\bk} (\rho)\,, 
\end{eqnarray}
with
\begin{eqnarray}
 \psi_{\bk}(\rho) = v_k(\rho) a_{\bk} + v_k^*(\rho) a^\dagger_{-\bk}\,,
\end{eqnarray}
where the creation and annihilation operators satisfy the commutation
relation 
$$[ a_{\sbm{k}},\, a^\dagger_{\sbm{p}}]=\delta^{(3)}(\bm{k}-\bm{p}).$$

In Ref.~\cite{IRgauge}, it was shown that 
the regularity of the one-loop corrections to
the two-point function implies that the positive frequency function
should satisfy
\begin{align}
 & \left[\left(1+\mu_2\right)\partial_\rho -(1+\lambda_2)
  X^i \partial_{X^i}+\mu_1 -2{\cal L}_k^{-1}\mu_3  \right] v_k(\rho)
 e^{i \bk \cdot \sbm{X}} = -(1+\lambda_2) D_k v_k(\rho)
 e^{i \bk \cdot \sbm{X}}\,, \label{Exp:GI}
\end{align}
where $D_k$ and ${\cal L}_k$ are defined as
\begin{equation}
 D_k := \bm{k}\cdot \partial_{\sbm{k}} + 3/2, \qquad
{\cal L}_k :=  \partial_\rho^2 +  (3 - \varepsilon_1+ \varepsilon_2) \partial_\rho
 + \frac{e^{-2\rho}}{{\dot\rho}^2} k^2\,,
\end{equation}
and we introduced the time dependent functions $\mu_i\,(i=1,\,2,\,3)$
written in terms of the horizon-flow functions as
\begin{align}
 \mu_1:= \varepsilon_1 + \frac{1}{2} \varepsilon_2 + 2 \xi_2\,,\qquad
 \mu_2:=  \varepsilon_1 \left( 1 + \varepsilon_1 + \varepsilon_2
 \right) + \lambda_2\,,\qquad
 \mu_3:= \frac{3}{4}\varepsilon_2 (2\varepsilon_1+\varepsilon_2)\,.
 \label{Def:mui}
\end{align}
The condition (\ref{Exp:GI}) should be satisfied for all wavelengths and
restricts the behaviours of not only the IR modes but also the UV
modes. 
This condition can be shown to be
consistent with the mode equation (\ref{Eq:zetan1}).    
At the leading order in the slow-roll approximation, this
condition, together with the usual condition on the UV behaviour,
completely determines the initial state to the Bunch-Davis vacuum. 
However, if we extend our argument to the higher order in the 
slow roll expansion, Eq.~(\ref{Exp:GI}) gives a non-trivial condition on 
the initial state. It might be
surprising that such a gauge-invariance condition (= IR regularity 
condition) significantly constrains the allowable initial state.

A gauge invariance condition on the coefficients 
of the homogeneous solutions is again derived from the regularity
conditions, although it is not sufficient to completely fix 
both $\xi_2$ and $\lambda_2$. 
Imposing this gauge invariance condition, the 
potentially divergent terms in loop corrections add up to be 
total derivative terms with respect to $k$ and hence they cease to 
contribute to the IR divergences. 
As discussed in Ref.~\cite{IRgauge}, the time-dependent functions
$\xi_2$ and $\lambda_2$ should also satisfy another relation to
keep the commutation relation between $\zeta_n$ and its
conjugate momentum satisfied. 
Assuming that $\xi_2$ and
$\lambda_2$ are appropriately determined, we do not discuss this
relation any further because their explicit forms are not important for
our current discussion.

\section{Primordial non-Gaussianities}  \label{Sec:NG}
In this section, we evaluate the primordial non-Gaussianity in the 
geodesic normal coordinates for the 
vacuum that satisfies the gauge-invariance conditions 
described in the preceding section. 
Our main focus is on the so-called consistency relation that 
the bispectrum in 
the squeezed limit is given by the amplitude 
and the spectral index of the power spectrum. 
We will show that, 
if we consider the genuine gauge-invariant correlation 
functions as defined in the preceding section, 
the leading term in the squeezed limit 
that arises in the usual computation of the bispectrum 
vanishes, and the well-known consistency relation 
does not hold.

\subsection{Standard consistency relation}
Now, we quickly review the derivation of the usual consistency relations 
for tree-level bispectra 
$ \langle \zeta_{\sbm{k}_1} \zeta_{\sbm{k}_2} \zeta_{\sbm{k}_3}
\rangle$ and 
$  \langle \delta \gamma^s_{\sbm{k}_1} \zeta_{\sbm{k}_2}
\zeta_{\sbm{k}_3} \rangle$. 
We are interested in the squeezed limit $k_1 \ll k_2 \simeq k_3$, 
where how we treat long wavelength modes becomes most essential.  
In this case 
the scale corresponding to $k_1$ crosses the Hubble horizon
much earlier than the others, so that 
$\zeta_{\sbm{k}_1}$ and $\delta \gamma_{\sbm{k}_1}$
become approximately constant by 
the time when the scales corresponding to $k_2$ and $k_3$ cross the horizon.
Therefore the mode with $k_1$ affects the evolution of the other modes
mostly through deformation of the background geometry. 
The main effect of this deformation 
is taken into account by shifting the wavenumbers as
\begin{align}
 & {\bm k}_a \, \to \quad \tilde{\bm k}_a = e^{- \zeta_{\sbm{k}_1}} \left[ e^{-
 \frac{1}{2}\delta \gamma_{\sbm{k}_1}} \right]\, {\bm k}_a=
 {\bm k}_a  - \left(\zeta_{\sbm{k}_1} +  {1\over 2} \delta
\gamma_{\sbm{k}_1} \right) {\bm k}_a + \cdots\,,  
\end{align}
where $a=2,\,3$. This modification influences the 
fluctuations $\zeta_{\sbm{k}_2}$ and $\zeta_{\sbm{k}_3}$, leading to 
the replacement
\begin{align}
 & \zeta_{{\sbm{k}}_a} 
\to (\det h)_{{\sbm k}_1}^{-1/2}
   \zeta_{\tilde{\sbm{k}}_a} 
 =  \zeta_{\sbm{k}_a} 
- \zeta_{\sbm{k}_1} \left( {\bm k}_a \!\cdot\!  \partial_{{\sbm k}_a} + \frac{3}{2}  \right)
 \zeta_{\sbm{k}_a} 
 - \frac{1}{2} {\bm k}_a \!\cdot\! \delta \gamma_{\sbm{k}_1} 
  \!\cdot\! \partial_{{\sbm k}_a} \zeta_{{\sbm k}_a}+  \cdots \,, \label{Exp:zetasqueeze}
\end{align}
where the factor $(\det h)_{{\sbm k}_1}^{-1/2}$ takes care of 
the change of the volume due to the modes with $\bm{k}_1$. 
Using the relation (\ref{Exp:zetasqueeze}), we can easily
evaluate the bispectrum for $\zeta$ in the squeezed limit as
\begin{eqnarray}
 \langle \zeta_{\sbm{k}_1} \zeta_{\sbm{k}_2} \zeta_{\sbm{k}_3} \rangle
  & \simeq &  
  - P_{k_1} 
\left( \partial_{\log k_2} +\partial_{\log k_3} 
+ 3 \right)  \langle
 \zeta_{\sbm{k}_2}\, \zeta_{\sbm{k}_3} \rangle\cr 
  & \simeq &  
  - P_{k_1} \delta^{(3)}\! \left(\bm{k}_2 +\bm{k}_3 \right)
  \left[ \partial_{\log k_2} + 3 \right] P_{k_2} 
 \simeq - \delta^{(3)} \left(\sum_{a=1}^3 \bm{k}_a \right)
 (n_s - 1) P_{k_1} P_{k_2} ~, \label{Exp:3pointzeta}
\end{eqnarray}
where $P_k$ is the unnormalized amplitude of the two-point function for
$\zeta_{\sbm{k}}$ defined by
\begin{eqnarray}
 \langle \zeta_{\sbm{k}_1} \zeta_{\sbm{k}_2} \rangle
 = \delta^{(3)} (\bm{k}_1 + \bm{k}_2) P_{k_1}\,, 
\end{eqnarray}
and $n_s$ is the spectral index. 
Thus, one can see that the three-point function in the squeezed limit 
is represented by $n_s$ and $P_k$.  
Equation~(\ref{Exp:3pointzeta}) is
called the consistency relation in single-field 
models of inflation.    

In a similar way, expanding the gravitational waves as
\begin{align}
 & \delta \gamma_{ij, \sbm{k}} = \sum_{s=\pm} e_{ij}^s(\bm{k}) \delta
 \gamma_{\sbm{k}}^s\,,
\end{align}
with the polarization tensors defined by 
\begin{align}
 {e^{s,i}}_{i}(\bm{k}) = 0 =  k^i e^s_{ij}(\bm{k})\,,  
\end{align}
the three-point function for $\delta\gamma\zeta\zeta$ 
is given by 
\begin{align}
  \langle \delta \gamma^s_{\sbm{k}_1} \zeta_{\sbm{k}_2} \zeta_{\sbm{k}_3} \rangle
  &\simeq - \frac{1}{2} \langle \delta \gamma^s_{\sbm{k}_1} \delta
 \gamma^s_{-\sbm{k}_1} \rangle \left(
\bm{k}_2\!\cdot\! \bm{e}^s(-\bm{k}_1)\!\cdot\!\partial_{\sbm{k}_2} 
+\bm{k}_3\!\cdot\! \bm{e}^s(-\bm{k}_1)\!\cdot\!\partial_{\sbm{k}_3} 
  \right)
 \langle \zeta_{\sbm{k}_2}\, \zeta_{\sbm{k}_3} \rangle \cr &
\simeq  \frac{4-n_s}{2} \delta^{(3)} 
  \left(\sum_{a=1}^3 \bm{k}_a \right)   \frac{\bm{k}_2\!\cdot\!
 \bm{e}^s(-\bm{k}_1)\!\cdot\! \bm{k}_2
}{k_2^2} P^{(\delta \gamma)}_{{k}_1} P_{{k}_2}\,, \label{Exp:3pointzetaGW}
\end{align}
where $P^{(\delta \gamma)}_k$ is the amplitude of the two-point function for the gravitational waves. These three-point functions
(\ref{Exp:3pointzeta}) and (\ref{Exp:3pointzetaGW}) were first
calculated by Maldacena in Ref.~\cite{Maldacena} and these results were extended to more general
single field models in Ref.~\cite{Creminelli:2004yq}.

\subsection{Gauge-invariant curvature perturbation} \label{SSec:GI}

In the following,  
we would like to consider the tree-level three-point function
for genuine gauge-invariant variables. 
To compare the results with the standard computation described 
in the preceding subsection, 
it would be better to use $\zeta$ instead of $R$ as a variable, 
but $\zeta$ itself 
does not transform as a scalar on a $t$-constant hypersurface. 
Therefore the curvature perturbation evaluated in the geodesic normal coordinates:
\begin{align}
 \gz(X):= \zeta (\rho,\,x^i(\bm{X}))
=\zeta(X) + \delta x^i \partial_i \zeta |_{x^i=X^i}+\cdots\,
\label{Def:gz} 
\end{align}
is not a genuine gauge invariant variable.  
However, this quantity $\gz(X)$ is related to the genuine gauge invariant variable truncated at the
required order in calculating the tree-level bi-spectrum:
\begin{align}
 & {^g\!R} =\sR + \delta x^i \partial_i \sR =  -2e^{-2(\rho+\zeta)} 
  \left[ e^{-\delta \gamma} \right]^{ij} \left( 2 \partial^2_{ij} \zeta
 + \partial_i \zeta \partial_j \zeta \right)  - 
\delta x^i \partial_i (-4 e^{-2\rho}) \partial^2 \zeta +  \cdots,  
\end{align}
as 
\begin{align}
 & {^g\!R} = -2 e^{-2\rho} \left( 2 \left({\partial\over\partial \bm{X}}
  \right)^2 \gz + {\partial\over\partial \bm{X}} \gz
 {\partial\over \partial \bm{X}} \gz \right)  + \cdots\,. \label{Exp:gR}
\end{align}
Here we neglected the terms quadratic in gravitational wave perturbation. 
For the modes with $k \gg 1/L_{\rm obs}$, using the
Fourier decomposition of the variables,
spatial derivatives can be replaced with the wave number, 
and hence Eq.~(\ref{Exp:gR}) implies that
$\gz$ is equivalently gauge-invariant as ${^g\!R}$. 
Therefore, in the succeeding subsection  we consider 
the three-point function for $\gz$ in Fourier space.

Now, we generalize Eq.~(\ref{Exp:Xi}) to be consistent
with our situation $1/L_{\rm obs} \ll k \ll aH $. 
In spite of this effort, it turns out later that 
the deviation from the formula (\ref{Exp:Xi}) does not contribute 
to the leading order terms in the squeezed limit.
Below, we mainly consider Fourier components of three-point
function. One may wonder if the use of Fourier components is in
conflict with our philosophy that genuine gauge invariant
quantities must be defined only in terms of the information contained 
in our local observable universe. Here the relevant modes are restricted to sufficiently 
short wavelength mode in the sense that $k\gg1/L_{\rm obs}$. In this case, we 
can take Fourier components after multiplying a smooth window function.

Focusing on the contribution from 
a single mode with the wavenumber $\bm{k}$, 
the geodesic equation is approximated by 
\begin{eqnarray}
 {{\rm d}^2 x^i\over {\rm d} \lambda^2}
\approx -i\left[2(\Bk\cdot\BX)X^i-X^2 k^i\right]\zeta_{\sbm{k}}
e^{i \sbm{k} \cdot \sbm{X}}
 - i \sum_{s=\pm} \left[ (\Bk\cdot\BX)X^j \delta\gamma_{~j,
		   {\sbm{k}}}^{i}  -
 \frac{1}{2}k^i X^j X^k \delta\gamma_{jk, \sbm{k}}  \right] 
e^{i \sbm{k} \cdot \sbm{X}}\,,
\label{geodesiceq2}
\end{eqnarray}
where we neglected the non-linear terms, that
do not contribute to the tree-level bi-spectrum.
This equation is integrated form $\lambda=0$ to 1 with the initial
condition (\ref{IC}). 
The result of the integration 
(\ref{geodesiceq2}) would be given in the form  
\begin{equation}
x^i(\BX)\approx e^{-\zeta(\sbm{X})} \left[ e^{- \delta \gamma(\sbm{X})/2}
				   \right]\!^i\!_j X^j +\Delta
x^i[\psi_\bk, \delta \gamma_{\bk}](\BX) 
 +\delta x_L^i[\psi_{\sbm{k}}](\BX) + \delta x_T^i[\delta \gamma_{\sbm{k}}](\BX),
\label{xofX}
\end{equation}
where we introduced
\begin{align}
 & \Delta x^i[\psi_\bk, \delta \gamma_{\bk}](\BX) :=
 \left[\psi_\bk(\BX)-\psi_\bk(0)  \right] X^i + \left[ \delta
 \gamma_{\bk,ij}(\BX)- \delta \gamma_{\bk,ij}(0)  \right] X^j =: 
 \Delta x_L^i + \Delta x_T^i\,.
\end{align}
The third and fourth terms in (\ref{xofX})
are corrections due to the curvature perturbation and the
gravitational waves contained in the Christoffel symbol, respectively.
As long as the contribution from a 
single mode $\psi_{\sbm{k}}$ is concerned, 
$\delta x^i_L$ is given by 
\begin{eqnarray}
\delta x_L^i[\psi_\bk](\BX) & = & 
  -i \psi_{\sbm{k}}
\left[2(\Bk\cdot\BX)X^i-X^2 k^i\right]
\int_0^1 {\rm d} \lambda\int_0^\lambda {\rm d} \lambda'   e^{i\bk\cd \bX\lambda'}\cr
& = & - \psi_{\sbm{k}}
\left[2(\Bk\cdot\BX)X^i-X^2 k^i\right] {i \over k^2}
  \int_0^k {\rm d} k' \int_0^{k'} {\rm d} k'' 
  e^{i\bk''\cd\bX}\cr
& = & \psi_{\sbm{k}}\, {i \over k^2}
  \int_0^k {\rm d} k' \int_0^{k'} {\rm d} k'' 
 \left[ 2(\Bk\cdot\Bpartial_{\bk''})\partial_{\bk''}^i-k^i \partial_{\bk''}^2\right]
  e^{i\bk''\cd\bX}
=: \psi_{\sbm{k}}\, \hat{\cal K}^{(L)i}_{\bk,\bk''}
e^{i\bk''\cd\bX}, 
\end{eqnarray} 
where $\Bk''$ is to be understood as $k''\Bk/k$.
The tensor contribution $\delta x^i_T$ can be similarly given by
\begin{align}
 \delta x_T^i[\delta \gamma_{\bk}](\BX) 
  &= \sum_{s=\pm} \delta \gamma_{\bk}^s  {i \over k^2}
  \int_0^k {\rm d} k' \int_0^{k'} {\rm d} k'' 
 \left[ (\Bk\cdot\Bpartial_{\bk''})  e^{s,i}\!_j \partial_{\bk''}^j 
 - \frac{1}{2} k^i e^s_{jk} \partial^j_{\bk''} \partial^k_{\bk''} \right]
  e^{i\bk''\cd\bX} \cr &=:  \sum_{s=\pm} \delta \gamma_{\bk}^s\, \hat{\cal K}^{(T,\,s)i}_{\bk,\bk''}
e^{i\bk''\cd\bX}\,.
\end{align}

\subsection{Primordial non-Gaussianity measured in the geodesic normal coordinates}

In the present and succeeding subsections, 
we calculate the primordial non-Gaussianity measured in the
geodesic normal coordinates. 
First, 
we neglect $\Delta x$, $\delta x_L$ and $\delta x_T$
in Eq.~(\ref{xofX}). In this case, using Eq.~(\ref{Exp:zeta}), 
the gauge-invariant curvature perturbation is given by
\begin{align}
 \gz&\Approx \zeta_n + \frac{1}{4} \varepsilon_2 \zeta_n^2
  + \zeta_n (\partial_\rho - X^i \partial_{X^i}) \zeta_n
  - \frac{1}{2} \delta \gamma^{ij} X_j \partial_i \zeta_n\,. \label{Exp:gz2nd}
\end{align}
In the succeeding subsection we will extend our discussions 
to incorporate the effects of 
$\Delta x$, $\delta x_L$ and $\delta x_T$, 
where we will find that this extension does not 
alter our main conclusion. 
Substituting the expression of $\zeta_n$ given in 
Eq.~(\ref{Sol:zetan2}) into Eq.~(\ref{Exp:gz2nd}), we obtain 
\begin{align}
  \gz(X)& \Approx \psi + \frac{1}{2} \mu_1\psi^2 + \left(1 + \mu_2 \right)
 \psi \partial_\rho \psi - {\cal L}^{-1} \mu_3 \psi^2 -(1+\lambda_2)  
 \psi X^i \partial_{X^i} \psi\,,  \label{Eq:gzeta}
\end{align}
up to second order in perturbation, where the time-dependent coefficients
$\mu_i$ with $i=1,2,3$ are defined in Eqs.~(\ref{Def:mui}).
Now, it is obvious that the contributions from the gravitational waves
are cancelled in the genuine gauge-invariant curvature perturbation
$\gz$ and therefore the contraction of $\gz$ with 
$\delta \gamma_{ij}$ trivially vanishes. The second-order perturbation
in gravitational waves $\delta \gamma_{ij,2}$ also contain the terms 
with two $\psi$s that
can yield non-vanishing terms in the three-point function for 
$\gz \gz \delta \gamma$. However,
both of $\psi$s in $\delta \gamma_{ij,2}$
are associated with derivatives. We therefore arrive at the conclusion
that the leading terms in 
$\langle \zeta \zeta \delta \gamma \rangle$ obtained in
Eq.~(\ref{Exp:3pointzetaGW}) do not remain, 
once we use $\gz$ in stead of $\zeta$. 

Next we calculate the bispectrum of $\gz$s. 
Introducing the Fourier modes of $\gz$ as
\begin{align}
 &\gz_{\sbm{k}}(\rho) = \int \frac{{\rm d}^3 \bm{X}}{(2\pi)^{3/2}}\, 
 e^{- i \sbm{k} \cdot \sbm{X}} \gz(\rho,\, \bm{X})\,, 
\end{align}
we consider $\langle \gz_{\bk_1} \gz_{\bk_2} \gz_{\bk_3} \rangle $ 
at the leading order in perturbation.
Expanding $\gz$ as $\gz= \psi + \gz_2+ \cdots$, we have
\begin{eqnarray}
 \langle \gz_{\bk_1} \gz_{\bk_2} \gz_{\bk_3} \rangle = 
  \langle \psi_{\bk_1} \psi_{\bk_2} \gz_{\bk_3,2}  \rangle + 
  \langle \psi_{\bk_1}  \gz_{\bk_2,2} \psi_{\bk_3}  \rangle +
   \langle  \gz_{\bk_1,2} \psi_{\bk_2} \psi_{\bk_3}  \rangle \label{Exp:Fg}
\end{eqnarray}
Here the last term in Eq.~(\ref{Exp:Fg}) does not contribute 
to the leading term in the squeezed limit, $k_1\ll k_2, k_3$ because the power spectrum for the mode with $k_1$ does not appear.

Using Eq.~(\ref{Eq:gzeta}), the first term on the right-hand side of
Eq.~(\ref{Exp:Fg}) is recast into 
\begin{align*}
  \langle \psi_{\bk_1} \psi_{\bk_2}  \gz_{\bk_3, 2} \rangle  
 = \left\langle \psi_{\bk_1} \psi_{\bk_2}  \left\{  (1+ \mu_2) \left( \psi \partial_\rho \psi
 \right)_{\bk_3}- (1+\lambda_2) \left( \psi X^i \partial_{X^i}
 \psi \right)_{\bk_3}\! +  \frac{\mu_1}{2}\! \left(\psi^2
 \right)_{\bk_3}\! \right\}\! \right\rangle 
 - \langle  \psi_{\bk_1} \psi_{\bk_2} {\cal L}^{-1}_{k_3} \mu_3 
 \left(\psi^2 \right)_{\bk_3} \rangle\,,  
\end{align*}
where we defined 
\begin{eqnarray}
 \left(\psi O \psi \right)_{\sbm{k}} 
:= \int \frac{{\rm d}^3
  {\bm X}}{(2\pi)^{3/2}} e^{-i \sbm{k} \cdot {\sbm X}} 
   \left( \prod_{i=1,2} \int \frac{{\rm d}^3 {\bm p}_i}{(2 \pi)^{3/2}}
   \right) e^{i \sbm{p}_1 \cdot \sbm{X}} \psi_{\sbm{p}_1} O
     e^{i \sbm{p}_2 \cdot \sbm{X}} \psi_{\sbm{p}_2}\,,
\end{eqnarray}
with $O=1,\, \partial_\rho,\, X^i \partial_{X^i}$ to concisely 
denote the terms originating from $\zeta_{n,2}$. The interaction 
picture fields $\psi_{\sbm{k}_1}$ and $\psi_{\sbm{k}_2}$ should
be contracted with either $\psi_{\sbm{p}_1}$ or $\psi_{\sbm{p}_2}$ 
contained in $ \left(\psi O \psi \right)_{\sbm{k}}$. 
Using the commutation relation for $a_{\sbm{k}}$ and 
$a^\dagger_{\sbm{k}}$, the first term of Eq.~(\ref{Exp:Fg}) is 
rewritten as
\begin{align}
 &\langle \psi_{\bk_1} \psi_{\bk_2}  \gz_{\bk_3,2} \rangle  \cr
 &\quad = v_{k_1} v_{k_2}\!\! \int \!\! \frac{\dd^3 \bm{p}}{(2\pi)^{3/2}}
 \int \frac{\dd^3 \bm{X}}{(2\pi)^3} \biggl[ e^{- i (\bk_1 + \bk_3) \cdot
 \sbm{X}} \delta^{(3)}(\bm{k}_2 + \bm{p}) v_{k_1}^*\! \left\{
 (1+\mu_2)\partial_\rho - (1+\lambda_2)  X^i \partial_{X^i}  + \mu_1
 -2 {\cal L}_p^{-1} \mu_3 \right\} v_p^* e^{i \bp \cdot \sbm{X}} \cr
 & \qquad \qquad \qquad \qquad \qquad \qquad \qquad -2 e^{-
 i \sbm{K} \cdot \sbm{X}} \delta^{(3)}(\bm{k}_2 + \bm{p}) \left\{
  {\cal L}_{k_3}^{-1}\mu_3 v_{k_1}^* v_{k_2}^* -  v_{k_1}^* {\cal
 L}_{k_2}^{-1}\mu_3 v_{k_2}^* \right\} \cr
 & \qquad \qquad \qquad \qquad \qquad \qquad \qquad + e^{-
 i(\bk_2+\bk_3)\cdot \sbm{X}}  \delta^{(3)}(\bm{k}_1 + \bm{p}) v_{k_2}^*
 \left\{(1+\mu_2)\partial_\rho - (1+\lambda_2) X^i
 \partial_{X^i} \right\} v_p^*  e^{i \bp \cdot \sbm{X}}  \biggr]\,,
  \label{Eq:3pointpart}
\end{align}
where we defined $\bm{K}:= \bm{k}_1+\bm{k}_2+\bm{k}_3$. Here, we
inserted the last terms on the first and second lines, which 
cancel with each other.

Following a similar calculation, the second term of Eq.~(\ref{Exp:Fg})
is rewritten as 
\begin{align}
 &\langle \psi_{\bk_1} \gz_{\bk_2,2}  \psi_{\bk_3} \rangle  \cr
 &\quad = v_{k_1} v_{k_3}^*\!\! \int \!\! \frac{\dd^3 \bm{p}}{(2\pi)^{3/2}}
 \int \frac{\dd^3 \bm{X}}{(2\pi)^3} \biggl[ e^{- i (\bk_1 + \bk_2) \cdot
 \sbm{X}} \delta^{(3)}(\bm{k}_3 + \bm{p}) v_{k_1}^*\! \left\{
 (1+\mu_2)\partial_\rho - (1+\lambda_2)  X^i \partial_{X^i}  + \mu_1
 -2 {\cal L}_p^{-1} \mu_3 \right\} v_p e^{i \bp \cdot \sbm{X}} \cr
 & \qquad \qquad \qquad \qquad \qquad \qquad \qquad -2 e^{-
 i \sbm{K} \cdot \sbm{X}} \delta^{(3)}(\bm{k}_3 + \bm{p}) \left\{
  {\cal L}_{k_2}^{-1}\mu_3 v_{k_1}^* v_{k_3} -  v_{k_1}^* {\cal
 L}_{k_3}^{-1}\mu_3 v_{k_3} \right\} \cr
 & \qquad \qquad \qquad \qquad \qquad \qquad \qquad + e^{-
 i(\bk_2+\bk_3)\cdot \sbm{X}}  \delta^{(3)}(\bm{k}_1 + \bm{p}) v_{k_3}
 \left\{(1+\mu_2)\partial_\rho - (1+\lambda_2) X^i
 \partial_{X^i} \right\} v_p^*  e^{i \bp \cdot \sbm{X}}  \biggr]\,.
  \label{Eq:tFgsecond}
\end{align}
Replacing $\bm{X}$ with $-\bm{X}$,    
we find that this expression is identical to the complex conjugate 
of Eq.~(\ref{Eq:3pointpart}) with the exchange between $\bm{k}_2$ and
$\bm{k}_3$, which means 
\begin{align}
 &  \langle \psi_{\bk_1}  \gz_{\bk_2, 2} \psi_{\bk_3}
 \rangle = \langle \psi_{\bk_1}  \psi_{\bk_3} \gz_{\bk_2, 2}
 \rangle^*\,. \label{Rel:1st2nd}
\end{align}
Here we noted that at superhorizon scales $v_{k_1}$ becomes constant and commutes
with the time integrals in ${\cal L}_{k_i}^{-1}$ where $i=2,3$.
It is, therefore, sufficient to calculate the first term of
Eq.~(\ref{Exp:Fg}). Using the gauge-invariance condition (\ref{Exp:GI}),
the first line in Eq.~(\ref{Eq:3pointpart}) can be 
recast into the simple expression:\footnote{One remark is in order about the translational invariance. There arose 
many terms that are not accompanied by the delta function $\delta^{(3)}(\BK)$. 
Usually appearance of the factor $\delta^{(3)}(\BK)$ 
is guaranteed by the momentum conservation. 
However, the origin of the spatial coordinates is fixed in the
geodesic normal coordinates we introduced, which manifestly breaks 
the global translational invariance. Therefore we cannot expect that 
the final result should be proportional to $\delta^{(3)}(\BK)$. 
} 
\begin{align}
&\mbox{(First line in Eq.~(\ref{Eq:3pointpart}))}\cr
 &\, = -(1+\lambda_2) 
  v_{k_1} v_{k_2}\!\! \int \!\! \frac{\dd^3 \bm{p}}{(2\pi)^{3/2}}
 \int \frac{\dd^3 \bm{X}}{(2\pi)^3} 
 v_{k_1}^*  e^{- i (\bk_1 + \bk_3) \cdot
 \sbm{X}} \delta^{(3)}(\bm{k}_2 + \bm{p})  D_p (v_p^*
 e^{i\sbm{p}\cdot\sbm{X}})
\cr
 & =- (1+\lambda_2) |v_{k_1}|^2 v_{k_2}\!\! \int \!\! \frac{\dd^3 \bm{p}}{(2\pi)^{3/2}}
   \delta^{(3)}(\bm{k}_2+\bm{p})  \left( \bm{p} \cdot \partial_{\sbm{p}}
 + 3/2 \right) 
   \left(v_p^*\delta^{(3)}(\bm{k}_1+\bm{k}_3-\bm{p}) \right)
\cr
 &\,= - \frac{1}{2} (1+\lambda_2) |v_{k_1}|^2 v_{k_2}
 v^*_{|\bk_1+\bk_3|} \!\! \int \!\! \frac{\dd^3 \bm{p}}{(2\pi)^{3/2}} \biggl[
 \delta^{(3)}(\bm{k}_2+\bm{p})  \left( \bm{p} \cdot \partial_{\sbm{p}}  + 3/2 \right) 
   \delta^{(3)}(\bm{k}_1+\bm{k}_3-\bm{p}) 
\cr
 & \qquad \qquad \qquad \qquad\qquad \qquad  \qquad\qquad \qquad \qquad + \delta^{(3)}(\bm{k}_1+\bm{k}_3-\bm{p})
 \left\{(\bm{p}+\bm{k}_1-\bm{k}_2+\bm{k}_3) \cdot \partial_{\sbm{p}}  +
 3/2 \right\}  \delta^{(3)}(\bm{k}_2+\bm{p}) \biggr]
\cr&\,=  - \frac{1}{2} (1+\lambda_2) |v_{k_1}|^2 
v_{k_2} v^*_{|\bk_1+\bk_3|}
\biggl[(2\pi)^{-3/2}(\Bk_1-\Bk_2+\Bk_3)\!\cdot\!
 \partial_{\bK}\delta^{(3)}(\BK)  \cr
 & \qquad \qquad \qquad \qquad\qquad \qquad  \qquad\qquad \qquad \qquad
+ \!\! \int \!\! \frac{\dd^3 \bm{p}}{(2\pi)^{3/2}}\,\partial_{\bp}\!\cdot\!\Bp
  \, \delta^{(3)}(\Bp+\Bk_1+\Bk_3)\delta^{(3)}(\Bp-\Bk_2)
\biggr]\,. \label{Exp:3point1st}
\end{align}
In the third equality, we 
replaced $\Bp$ with $-(\Bp-\Bk_1+\Bk_2-\Bk_3)$
in the half of the expression. The second term in the last line takes
the form of total derivative and hence it does not contribute. 
The first term proportional to
$\bm{k}_1\cdot \partial_{\bK}\delta^{(3)}(\BK)$ in 
Eq.~(\ref{Exp:3point1st}) vanishes in the limit $k_1\to 0$. 
The other terms on the same line are not suppressed at this moment.
Combining the contribution from the second term of Eq.~(\ref{Exp:Fg}),
which satisfies Eq.~(\ref{Rel:1st2nd}), these terms, however, provide 
a factor $(v_{k_2} v^*_{|\bk_1+\bk_3|}-v_{|\bk_1+\bk_2|}v_{k_3}^*)$, 
which again vanishes in the squeezed limit, $k_1\to 0$.

To evaluate the second line in Eq.~(\ref{Eq:3pointpart}), 
we again note that $v_{k_1}$ commutes with ${\cal L}^{-1}_{k_3}$ for
$k_1 \ll e^{\rho} H$. Thus, we have
\begin{align}
&\mbox{(Second line in Eq.~(\ref{Eq:3pointpart}))}
 =-2 |v_{k_1}|^2 v_{k_2} (2\pi)^{-3/2}
 \delta^{(3)}(\bm{K})  \left\{
  {\cal L}_{|\sbm{k}_2+\sbm{k}_1|}^{-1}\mu_3 v_{k_2}^* -   {\cal
 L}_{k_2}^{-1}\mu_3 v_{k_2}^* \right\}\,, 
\end{align}
which vanishes in the limit $k_1\to 0$. 

As for the last line 
in Eq.~(\ref{Eq:3pointpart}),  
we notice that 
$\bm{p}$ can be replaced with $-\bm{k}_1$ 
owing to the factor $\delta^{(3)}(\bm{k}_1+\bm{p})$. 
Therefore the term 
with the $\rho$-derivative acting on $v^*_p$ is suppressed 
in the limit $k_1\to 0$. 
Hence, the terms in the third line can be
simplified to
\begin{align}
\mbox{(Last line in Eq.~(\ref{Eq:3pointpart}))}
 & =- (1+\lambda_2) |v_{k_1}|^2 |v_{k_2}|^2\!\! \int \!\! \frac{\dd^3
 \bm{p}}{(2\pi)^{3/2}} \delta^{(3)}(\bm{k}_1+\bm{p}) \bm{p} \cdot \partial_{\bp} 
 \delta^{(3)}(\bm{k}_2+\bm{k}_3-\bm{p}) 
 \cr &
 =  - (1+\lambda_2) |v_{k_1}|^2 |v_{k_2}|^2 (2\pi)^{-3/2} \bm{k}_1
 \cdot \partial_{\sbm{K}} \delta^{(3)}(\bm{K})\,. \label{Exp:3point2nd}
\end{align} 
This expression vanishes in the limit $k_1\to 0$.  
Thus we find that all the leading contributions in the tree-level
bispectrum Eq.~(\ref{Exp:Fg}) vanish in the squeezed limit.

\subsection{Remaining contributions from the transformation to geodesic normal coordinates} 
\label{GNC} 
Here we take into account the effects of 
$\Delta x$, $\delta x_L$ and $\delta x_T$. 
We calculate the three-point functions, whose three momenta are given by $\bm{p}_1$,
$\bm{p}_2$,\, and $\bm{p}_3$ without specifying which 
one is $\bm{k}_1$. 
We first discuss the scalar contribution $\Delta x^i_L$ to the
three-point function with cubic $\zeta$s, given by
\begin{eqnarray}
\int {{\rm d}^3\! X\over
 (2\pi)^{3/2}}e^{-i\bp_3\cd\bX}\langle\psi_{\bp_1} \Delta x_L^i\rangle
  \langle \psi_{\bp_2}\partial_i \psi\rangle
& = &
\int {{\rm d}^3\! X\over
(2\pi)^{3/2}}e^{-i\bp_3\cd\bX}\langle\psi_{\bp_1} \Delta x_L^i[\psi_{-\bp_1}](\BX)\rangle
  \langle \psi_{\bp_2}\partial_i \psi_{-\bp_2}\rangle\cr
& = & |v_{p_1}|^2|v_{p_2}|^2
\int {{\rm d}^3\! X\over
(2\pi)^{3/2}}e^{-i\bp_3\cd\bX}\left(e^{-i\bp_1\cd\bX}-1\right)X^i\partial_i
e^{-i\bp_2\cd\bX}
\cr & = & |v_{p_1}|^2|v_{p_2}|^2
\Bp_2\cd\Bpartial_{\bp_3}
\int {{\rm d}^3\! X\over
(2\pi)^{3/2}}e^{-i\bp_3\cd\bX}\left(e^{-i\bp_1\cd\bX}-1\right)e^{-i\bp_2\cd\bX}
\cr & = & |v_{p_1}|^2|v_{p_2}|^2 \Bp_2\cd\Bpartial_{\bp_3}
\left[\delta^{(3)}(\Bp_1+\Bp_2+\Bp_3)-\delta^{(3)}(\Bp_2+\Bp_3)\right].
 \label{Exp:DxLi}
\end{eqnarray}
When $\Bp_1$ is identified with $\Bk_1$, the factor
$\left[\delta^{(3)}(\Bp_1+\Bp_2+\Bp_3)-\delta^{(3)}(\Bp_2+\Bp_3)\right]$
vanishes in the limit $k_1\to 0$. 
When $\Bp_2$ is identified with $\Bk_1$, there is a manifest suppression 
factor proportional to $\Bp_2$. 
When $\Bp_3$ is identified with $\Bk_1$, 
the power spectrum $|v_{k_1}|^2$ is absent and Eq.~(\ref{Exp:DxLi}) does
not yield the dominant contribution in the squeezed limit. 
Therefore, for all cases, the contribution from $\Delta \Bx_L $
vanishes in the limit $k_1\to 0$. A similar argument follows for
the non-vanishing contributions of $\Delta \Bx_T$ to the three-point
functions.

Next, we consider the contributions from $\delta\Bx_L$ and $\delta \Bx_T$.
Again we evaluate a rather general expression: 
\begin{eqnarray}
\int {{\rm d}^3\! X\over (2\pi)^3}e^{-i\bp_3\cd\bX}\langle\psi_{\bp_1}\delta x_L^i\rangle
  \langle \psi_{\bp_2}\partial_i \psi\rangle
&=&
|v_{p_1}|^2|v_{p_2}|^2
\int {{\rm d}^3\! X\over (2\pi)^3}e^{-i\bp_3\cd\bX}\hat{\cal K}^{(L)i}_{-\bp_1,\bk''}
e^{i\bk''\cd\bX} \partial_i e^{-i\bp_2\cd\bX}
\cr &=&
-|v_{p_1}|^2|v_{p_2}|^2
 {p_{2,i}\over p_1^2} 
  \int_0^{p_1}\!\! {\rm d} k' \int_0^{k'}\!\! {\rm d} k'' 
 \left[ 2(\Bp_1\cdot\Bpartial_{\bk''})\partial_{\bk''}^i-p_1^i
  \partial_{\bk''}^2\right] \delta^{(3)}(\Bp_2+\Bp_3-\Bk''), \nonumber \\
\end{eqnarray}
where $\bm{k}''$ should be understood as $-k'' \bm{p}_1/p_1$.
Suppression is manifest in the case with $\bm{p}_2 = \bm{k}_1$ or 
$\bm{p}_3 = \bm{k}_1$. In the former case 
there exists a manifest
suppression factor proportional to $\Bp_2$, while in the later case 
there is no contribution from the spectrum $|v_{k_1}|^2$ as
before. We therefore examine the case in which $\Bp_1$ is identified with
$\Bk_1$. In this case, it will be convenient to rewrite the last
expression as  
\begin{align}
 &\int {{\rm d}^3\! X\over (2\pi)^3}e^{-i\bp_3\cd\bX}\langle\psi_{\bp_1}\delta x^i\rangle
  \langle \psi_{\bp_2}\partial_i \psi\rangle \cr
 &\qquad =-|v_{\bp_1}|^2|v_{\bp_2}|^2
 {p_{2,i}\over p_1^2} 
 \left[2(\Bp_1\!\cdot\!\Bpartial_{\bp_2})\partial_{\bp_3}^i-p_1^i
  (\Bpartial_{\bp_2}\!\cdot\!\Bpartial_{\bp_3})\right] 
  \int_0^{p_1}\!\! {\rm d} k' \int_0^{k'}\!\! {\rm d} k'' \delta^{(3)}(\Bp_2+\Bp_3-\Bk'')\,.
\end{align}
Denoting $\Bp_2+\Bp_3$ by $\Bp$, the part related to $\Bp_1$ is then simply given in the form, 
\begin{eqnarray}
{|v_{p_1}|^2\over p_1}g(\hat\Bp_1) 
  \int_0^{p_1}\!\! {\rm d} k' \int_0^{k'}\!\! {\rm d} k'' \delta^{(3)}(\Bp-\Bk'')
 & = & 
 {|v_{p_1}|^2\over p_1}g(\hat\Bp_1) 
  \int_0^{p_1}\!\! {\rm d} k' \int_0^{k'}\!\! {\rm d} k'' {1\over p^2}\delta(p-k'')
   \delta^{(2)}(\hat\Bp+\hat\Bp_1)\cr
 & = & 
 {|v_{p_1}|^2}
 {(p_1-p)^2 \over p_1p^2} \frac{\theta(p_1-p)}{p_1-p}  g(\hat\Bp_1) \delta^{(2)}(\hat\Bp+\hat\Bp_1)\,, 
\end{eqnarray}
where $\hat\Bp:=\Bp/p$, $\hat\Bp_1:=\Bp_1/p_1$, and $g(\hat\Bp):= 2(\hat\Bp\!\cdot\!\Bpartial_{\bp_2})\partial_{\bp_3}^i-\hat{p}^i
  (\Bpartial_{\bp_2}\!\!\cdot\!\Bpartial_{\bp_3})$.
If we integrate 
the above expression for a small region $p_1<\epsilon={\cal O}(p)$ 
multiplying $G(\Bp_1)$, which is an arbitrary smooth function of $\Bp_1$, 
it becomes ${\cal O}(\epsilon |v_{\bp}|^2 G(\bm{p}))$. The obtained expression is suppressed 
by a factor $\epsilon$ compared with the case when we consider an 
integral for the simple delta function, 
$\int {\rm d}^3 \bm{p}_1 G(\Bp_1){|v_{\sbm{p}_1}|^2}
\delta^{(3)}(\Bp+\Bp_1)$. Since this argument does not depend on the
details of the function $g(\hat{\Bp})$, a similar suppression can be
easily found also in the contribution from $\delta \Bx_T$. Now, we can
understand that all the neglected terms in the preceding subsection 
also vanish in the squeezed limit $k_1 \to 0$.

\subsection{Discussions}

The suppression in the squeezed limit does not 
occur in the usual computation of the bispectrum for $\zeta$. 
What we have found is that the leading term in the squeezed limit, 
exhibited in the consistency relation (\ref{Exp:3pointzeta}),
is significantly modified 
if we measure the gauge invariant curvature perturbation 
by using the geodesic normal coordinates.  
This can be understood from the fact that the non-Gaussianity 
described by the consistency relation is
originating from the deformation of the local coordinates due to the IR
modes that crossed the horizon much earlier than the other two modes.  
As shown in our previous works~\cite{IRgauge_L, IRgauge} and also in
Refs.~\cite{Giddings:2010nc, Byrnes:2010yc, Gerstenlauer:2011ti}, this
deformation can be absorbed by the gauge transformation in the local universe. This leads
to the conclusion that there should not be the net physical effect due to these IR
modes in this case. (See also
Ref.~\cite{Gerstenlauer:2011ti}, where a similar analysis is presented,
while the authors seem to have the slightly different opinion about what are
the actual observables.)

If we were thinking of the case with $k_1 \ll 1/L_{\rm obs}$, 
this cancellation can be understood rather easily as mentioned above. 
By contrast, our result might be a little surprising because the
longest wavelength mode $k_1$ is, in the present setup, supposed to be
still within the size of our observable universe. Nevertheless,  
the cancellation seems to continue to work. 
In the actual observation 
what we are interested in will be the temperature fluctuation 
of cosmic microwave background, which 
is not exactly what we computed here. 
In the case of CMB anisotropy, observation 
is performed not on a constant time hypersurface but 
along the light cone. Furthermore the observable 
region is not well outside the horizon even at the last 
scattering surface. Despite of these differences, 
our result suggests that for single field inflation models 
the leading terms of bi-spectrum in the squeezed limit 
may vanish also in the CMB anisotropy 
as long as we analyze the data by using 
observationally natural coordinates.

\section{Conclusion}  \label{Sec:Conclusion}
In this paper we reexamined the so-called consistency relation, 
focusing on what are genuine observables in our local universe. If we
knew the configuration of the whole universe, usual curvature perturbation $\zeta$ has gauge invariant meaning. 
However, the region we can observe is limited. 
In this case $\zeta$ itself is not genuine gauge invariant. 

In our previous paper we proposed a genuine gauge invariant variable 
that can be locally constructed, e.g. three curvature on $\phi$-constant 
hypersurface measured in the geodesic normal coordinates. 
We found that the two-point function of this quantity 
is free from IR divergence in single field inflation models 
only if the initial state satisfies 
what we called the gauge invariance conditions. 
Applying thus-obtained gauge invariance conditions, we investigated 
bispectrum in the squeezed limit, where one of the momenta
of the arguments of the three-point function is much less 
than the others. 

In the squeezed limit, the bispectrum is known to be
expressed by the power spectrum and the spectral index.
Although we are not claiming that the result of
this standard computation is wrong,
this conclusion is derived by evaluating the three-point function
for the so-called curvature perturbation, which is not
a genuine gauge invariant variable in the sense mentioned above.
We therefore re-investigated the behavior of the three-point
function for an alternative quantity, which is genuinely gauge invariant.
As a result, we found that the leading term in the squeezed limit
vanishes, i.e. the bispectrum is more suppressed in the
squeezed limit than we expect from the naive calculation based
on the so-called gauge-invariant curvature perturbation.

We addressed the local non-Gaussianity in single-field models with the
standard kinetic term, but our argument will be
easily extended to models with a non-standard kinetic
term. Since the large local non-Gaussianity is reported in multi-field
models, the extension to multi-field models are intriguing, too. For
this extension, we need to keep in mind the different nature of
the IR modes between in single-field models and in multi-field models~\cite{IRmulti}.
In the latter case, IR modes include physical degrees of freedom as well
as unphysical gauge degrees of freedom.
Therefore such perfect cancellation as we observed in single field models
will not be expected in a general multi-field case.
We will report these issues in our forthcoming publication.

\acknowledgments
The discussions during the workshops YITP-T-09-05, YITP-T-10-01, and
YKIS2010 at Yukawa Institute were very useful to complete this
work. T.~T. and Y.~U. would like to thank the hospitality of the Perimeter
Institute during the workshop ``IR Issues and Loops in de Sitter Space.''
Y.~U. would also like to thank Robert Brandenberger and Arthur Hebecker for the
hospitalities of McGill university and Heidelberg university. We also
acknowledge Robert Brandenberger, Chris Byrnes, Jaume Garriga, Mischa
Gerstenlauer, Arthur Hebecker, and Martin Sloth for the helpful discussions. T.~T. and Y.~U. are supported by
the JSPS under Contact Nos.\ 22840043 and 21244033. Y.~U. is also supported by
the JSPS under Contact No.\ 21244033, MEC FPA under Contact No.\ 2007-66665-C02, and MICINN
project FPA under Contact No.\ 2009-20807-C02-02. 
We also acknowledge the support
of the Grant-in-Aid for the Global COE Program ``The Next Generation of
Physics, Spun from Universality and Emergence'' and the Grant-in-Aid for
Scientific Research on Innovative Area Nos.\ 21111006 and 22111507 from the MEXT.

\end{document}